\documentclass{ws-ijqi}

\newcommand{\ket}[1]{\left\vert#1\right\rangle}

\newcommand{\minisand}[3]{\langle#1\vert#2\vert#3\rangle}

\begin{document}

\markboth{C. Di Franco, M. Paternostro, and G. M. Palma}
{A deeper insight into quantum state transfer from an information flux viewpoint}

\catchline{}{}{}{}{}

\title{A DEEPER INSIGHT INTO QUANTUM STATE TRANSFER FROM AN INFORMATION FLUX VIEWPOINT}

\author{C. DI FRANCO, M. PATERNOSTRO}

\address{School of Mathematics and Physics, Queen's University, Belfast BT7 1NN, United Kingdom\\
c.difranco@qub.ac.uk}

\author{G. M. PALMA}

\address{NEST-CNR (INFM) \& {D}ipartimento di Scienze Fisiche ed Astronomiche, Universita' degli studi di Palermo, via Archirafi 36, Palermo, 90123, Italy}

\maketitle


\vskip1cm

\begin{abstract}
We use the recently introduced concept of information flux in a many-body register in order to give an alternative viewpoint on quantum state transfer in linear chains of many spins.
\end{abstract}


\section{Introduction}

The control over multipartite devices, used for the purposes of quantum communication and quantum computation, can be reduced by means of specific form of built-in and permanent intra-register couplings. This possibility has recently been the center of a strong interest of the quantum information processing (QIP) community.\cite{general,bose,alwayson} The price to pay in order to avoid generally demanding fast and accurate inter-qubit switching and gating is the pre-engineering of appropriate patterns of couplings. Such a scenario is in general indicated as {\it control-limited}. The efforts are in this case directed towards the determination of the exact distribution of coupling strengths for a given inter-qubit interaction model in a control-limited setting. Here we discuss a recently proposed approach that sheds new light onto the achievement of this task and the design of efficient QIP protocols.

In Ref.~\refcite{informationflux} we have introduced the concept of {\it information flux} in a quantum mechanical system. This can be seen in terms of the influences that the dynamics of a selected element of a multipartite register experience due to {\it interaction channels} with any other parties. We showed that information can be effectively processed by arranging the network of interactions in a way so as to privilege or repress specific interaction channels. The realization of an optimal QIP task is therefore translated into the maximization of the information flux associated with such channels. Here we give a deeper insight into the mechanism behind this tool and, therefore, a clear and intuitive picture of it by  providing an explicit example.

This paper is organized as follows. In Sec.~\ref{concept} we resume the concept of information flux. In Sec.~\ref{transfer} we address the case of quantum state transfer in spin chains and describe a graphical method as an additional tool to perform such an analysis. Finally, we summarize our results in Sec.~\ref{remarks}.

\section{Concept}
\label{concept}
Let us consider a register of $N$ interacting qubits coupled via the Hamiltonian $\hat{\cal H}_{\{g\}}(t)$ whose structure we do not need to specify here. Our assumption is that $\hat{\cal H}_{\{g\}}(t)$ depends on a set of parameters $g_j$ (which could stand for the coupling strengths between the elements of the register) and a generalized time parameter $t$. We adopt the notation according to which $\hat{\Sigma}_j=\otimes^{j-1}_{k=1}\hat{I}_k\otimes\hat{\sigma}_{\Sigma_{j}}\otimes^{N}_{l=j+1}\hat{I}_{l}$ $(\Sigma=X,Y,Z)$ is the operator that applies the $\hat{\sigma}_{\Sigma}$ Pauli-matrix only to the $j$-th qubit of the register. Here $\hat{I}_{j}$ is the $2\times{2}$ identity matrix of qubit $j$. In the remainder of the paper, we work in the Heisenberg picture where time-evolved operators are indicated as $\hat{\tilde{O}}_j(t)=\hat{\cal U}^\dag(t)\hat{O}_j\hat{\cal U}(t)$ with $\hat{\cal U}(t)={\rm exp}[{-({i}/{\hbar})\int\hat{\cal H}_{\{g\}}(t')dt'}]$. 
\begin{figure}[t]
\center\psfig{figure=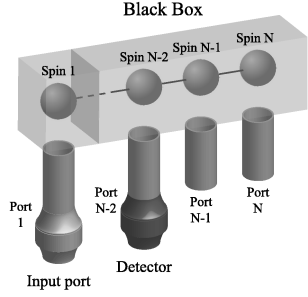,width=4cm}
\caption{Scheme of the processes we consider. A computation or communication step is interpreted as a black box (whose operation depends on the coupling scheme within a multipartite register of qubits) with movable input and a detection terminal. Through the study of information-flux dynamics, we can design the best coupling scheme for a chosen QIP operation.}
\label{fig:scheme}
\end{figure}
We say that {\it there is information \it extractable from qubit $j$ at time $t$} whenever there is at least one $\hat{\Sigma}$ for which $\langle\hat{\tilde\Sigma}_{j}(t)\rangle\neq{0}$. Here, the expectation value is calculated over the initial state of the register $\ket{\Psi_0}_{1..N}$.

We adopt, hereafter, the following schematic description of a computation or communication process: we suppose to have access to a selected qubit of a multipartite register and we consider it as the input terminal of the black box given by the rest of the elements and their mutual coupling. We then use a detection stage which can be attached to a suitable output port, connected to one of the qubits in the black box (a sketch is given in Fig.~\ref{fig:scheme}). In this picture, the initial state of a quantum system is described by the state vector $\ket{\Psi_0}_{1..N} = \ket{\phi_0}_1 \ket{\psi_0}_{2..N}$. This is the case in which the first qubit is initialized in a generic {\it input} state (and is separable with respect to the rest of the register) and $\ket{\psi_0}_{2..N}$ represents the initial state of the other qubits that, in general, can be mutually entangled. We assume this state to be known and independent of the input state. The assumption of a known initial state of the register $\{2,..,N\}$ corresponds to the situation assumed in many control-limited QIP protocols. We can thus interpret a quantum process as the {\it flux} of appropriately processed information from the input qubit to the remaining components of the register. Such a flux is witnessed by any explicit dependence of the dynamics of the $i$-th qubit on the operators associated with the input one. Therefore, in order to find if qubit $i$ has developed any extractable information at time $t$, as a result of an information flux from the input qubit, 
we need to study the dependence of $\minisand{\Psi_0}{\hat{\tilde\Sigma}_i(t)}{\Psi_0}$'s on at least one of $\minisand{\Psi_0}{\hat{\Sigma}'_1}{\Psi_0}$'s $(\Sigma,\Sigma'=X,Y,Z)$.

We refer to Ref.~\refcite{informationflux} for mathematical details and for the formal definition of the information flux. Here we would like just to specify that this approach consists in the decomposition of each $\hat{\tilde{\Sigma}}_i(t)$ over the operator-basis built out of all the possible tensorial products of single-qubit operators acting on the elements of the system $\{1,..,N\}$. The information flux can be evaluated from the expectation value (over $\ket{\psi_0}_{2..N}$) of the participants to this decomposition which include operators of the first qubit.

The control over the set \{g\} can be fully utilized in the preparation of a multipartite device in the most appropriate configuration of couplings and initial state, for a given QIP task. The potential of this approach is better illustrated in Refs.~\refcite{informationflux},~\refcite{matryoshka}. In what follows, in order to show a practical application of this method, we will address explicitly the case of quantum state transfer.

\section{Quantum State Transfer}
\label{transfer}

Quantum state transfer in spin chains is a scenario where the information flux approach is particularly useful. For short-distance communication, the idea of using spin chains as {\it quantum wires} has been put forward by Bose.\cite{bose} With an isotropic Heisenberg interaction and a local magnetic field, a transmission fidelity that exceeds the maximum value achievable classically can be obtained for a chain up to $\sim80$ qubits. Later, Christandl {\it et al.} showed that, by engineering the strength of the couplings in the chain, a unit fidelity can be reached in for end-to-end transfer in a linear chain, among topologically more complicated situations.\cite{cambridge}

We will first analyze an open three-qubit chain, whose Hamiltonian reads ${\cal\hat{H}}=\sum_{i=1}^2J(\hat{X}_i\hat{X}_{i+1}+\hat{Y}_i\hat{Y}_{i+1})$ with $J$ the coupling strength of the pairwise interaction between the qubits. This is an instance of the linear model considered in Ref.~\refcite{cambridge} and thus perfect state transfer from the first to the third qubit is obtained after a time $t^*=\pi/(2\sqrt{2}J)$ [we set $\hbar=1$ throughout the paper], if the initial state of spin 2 and 3 is $\ket{\psi_0}_{2,3}=\ket{00}_{2,3}$. If we want to determine the information flux from the first to the last qubit, we should evaluate the evolution of $\hat{X}_3$ and $\hat{Y}_3$ ($\hat{Z}_3$ can be seen as $-i\hat{X}_3\hat{Y}_3$) in the Heisenberg picture, at time $t$. It is easy to verify that
\begin{eqnarray}
  \nonumber&\hat{\tilde{X}}_3(t)=\alpha_1(t)\,\hat{X}_1\hat{Z}_2\hat{Z}_3+\alpha_2(t)\,\hat{Y}_2\hat{Z}_3+\alpha_3(t)\,\hat{X}_3,\\
  &\hat{\tilde{Y}}_3(t)=\beta_1(t)\,\hat{Y}_1\hat{Z}_2\hat{Z}_3+\beta_2(t)\,\hat{X}_2\hat{Z}_3+\beta_3(t)\,\hat{Y}_3.
\end{eqnarray}
with $\alpha_1(t)=\beta_1(t)= -\sin^2(\sqrt{2}Jt)$, $\alpha_2(t)=-\beta_2(t)=(1/\sqrt{2})\,\sin(2\sqrt{2}Jt)$, and $\alpha_3(t)=\beta_3(t)=\cos^2(\sqrt{2}Jt)$. The only term in the decomposition of the evolved operator $\hat{\tilde{X}}_3(t)$ ($\hat{\tilde{Y}}_3$(t)) in which $\hat{X}_1$ ($\hat{Y}_1$) is present is $\alpha_1(t)\,\hat{X}_1\hat{Z}_2\hat{Z}_3$ ($\beta_1(t)\,\hat{Y}_1\hat{Z}_2\hat{Z}_3$). The information flux from $\hat{X}_1$ to $\hat{X}_3$ ($\hat{Y}_1$ to $\hat{Y}_3$) is therefore ${\cal I}^{XX}_{3}(t)=\alpha_1(t)\,_{2,3}\minisand{00}{\hat{Z}_2\hat{Z}_3}{00}_{2,3}=\alpha_1(t)$ [${\cal I}^{YY}_{3}(t)=\beta_1(t)$]. For $t=t^*$, we have ${\cal I}^{XX}_{3}(t)={\cal I}^{YY}_{3}(t)=-1$. A perfect state transfer should correspond to an information flux between homonymous operators equal to 1. The minus sign we found can be explained by remembering that, in Christandl's protocol, we need to apply a single-qubit rotation 
\begin{equation}
\hat{R}(N)=
\begin{pmatrix} 
\:&1&\:\:&0\\ 
\:&0&\:\:&e^{\frac{i\pi(N-1)}{2}}
\end{pmatrix}
\end{equation}
on the last qubit, after the action of the Hamiltonian $\hat{{\cal H}}$, in order to compensate an additional phase factor arising from the evolution. The inclusion of this gate into our evolution just corresponds to change $\hat{X}_3$ in $-\hat{X}_3$ and $\hat{Y}_3$ in $-\hat{Y}_3$, keeping $\hat{Z}_3$ unchanged. We thus obtain information flux from $\hat{X}_1$ to $\hat{X}_3$ and from $\hat{Y}_1$ to $\hat{Y}_3$ (and obviously from $\hat{Z}_1$ to $\hat{Z}_3$) equal to $+1$, at time $t=t^*$.

In general, a decomposition over the operator-basis can be demanding, especially for a large number of qubits. Indeed, the dimension of this basis is $4^N$. In some cases, in virtue of the symmetries of the interaction model (as in the example analyzed above), the evolution of $\hat{\Sigma}_i$ involves only few elements of this basis. A simple method to estimate which terms are included in this evolution is presented here.

\begin{figure}[t]
\center\psfig{figure=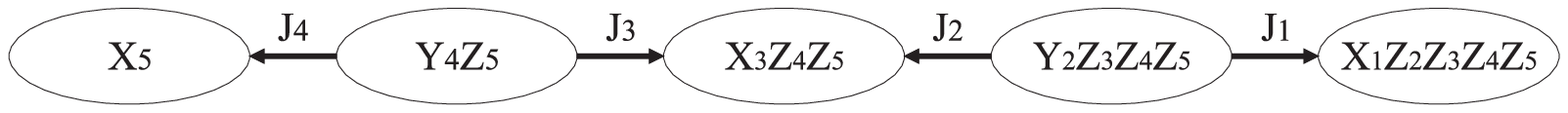,width=8cm}
\caption{Oriented graph summarizing the recurrence formulas to obtain the operator decomposition for a 5-qubit chain, whose Hamiltonian reads ${\cal\hat{H}}=\sum_{i=1}^4J_i(\hat{X}_i\hat{X}_{i+1}+\hat{Y}_i\hat{Y}_{i+1})$.}
\label{fig:graph}
\end{figure}

For a time-independent Hamiltonian, it is $\hat{\tilde{\Sigma}}_i(t)=e^{\frac{i}{\hbar}\hat{\cal H}t}\,\hat{\Sigma}_i\,e^{-\frac{i}{\hbar}\hat{\cal H}t}$ and by means of the operator expansion formula, we have
\begin{equation}
  \hat{\tilde{\Sigma}}_i(t)=\hat{\Sigma}_i+\frac{i}{\hbar}t[{\hat{\cal H}},\hat{{\Sigma}_i}]+\frac{1}{2!}(\frac{i}{\hbar}t)^2[{\hat{\cal H}},[{\hat{\cal H}},\hat{{\Sigma}_i}]]+..
\end{equation}
If the Hamiltonian $\hat{\cal H}$ is expressed in terms of operators $\hat{\Sigma}'_j$'s all the commutators can be easily represented in a graph. Suppose, for instance, that we want to analyze a 5-qubit chain, whose Hamiltonian reads ${\cal\hat{H}}=\sum_{i=1}^4J_i(\hat{X}_i\hat{X}_{i+1}+\hat{Y}_i\hat{Y}_{i+1})$, focusing on the evolution of $\hat{X}_5$. The first commutators are
\begin{equation}
 \begin{split}
  &[{\hat{\cal H}},\hat{X}_5]=-2i\,J_4\,\hat{Y}_4\hat{Z}_5,\\
  &[{\hat{\cal H}},\hat{Y}_4\hat{Z}_5]=2i\,J_3\,\hat{X}_3\hat{Z}_4\hat{Z}_5+2i\,J_4\,\hat{X}_5,\\
  &..
 \end{split}
\end{equation}
The only operators involved in this iterative sequence are $\hat{X}_5$, $\hat{Y}_4\hat{Z}_5$, $\hat{X}_3\hat{Z}_4\hat{Z}_5$, $\hat{Y}_2\hat{Z}_3\hat{Z}_4\hat{Z}_5$, and $\hat{X}_1\hat{Z}_2\hat{Z}_3\hat{Z}_4\hat{Z}_5$. Therefore it is possible to write the evolved operators $\hat{\tilde{X}}_5(t)$ as
\begin{eqnarray}
 \nonumber\hat{\tilde{X}}_5(t)=&\gamma_1(t)\hat{X}_1\hat{Z}_2\hat{Z}_3\hat{Z}_4\hat{Z}_5+\gamma_2(t)\hat{Y}_2\hat{Z}_3\hat{Z}_4\hat{Z}_5+\\
 &+\gamma_3(t)\hat{X}_3\hat{Z}_4\hat{Z}_5+\gamma_4(t)\hat{Y}_4\hat{Z}_5+\gamma_5(t)\hat{X}_5.
 \end{eqnarray}
In those cases where the parameters $\gamma_j(t)$ cannot be analytically evaluated due to the complications in the evolution, it is possible to approximate them by means of recurrence formulas. In our case, we have $\gamma_j(t)\sim\sum_{l=1}^M[(2t)^l/l!]\gamma_j^{(l)}$, where $M$ is a proper cut-off and
\begin{eqnarray}
 \nonumber&\gamma_1^{(l)}=-J_1\gamma_2^{(l-1)},\hskip2.5cm\gamma_2^{(l)}=J_1\gamma_1^{(l-1)}+J_2\gamma_3^{(l-1)},\\
 \nonumber&\gamma_3^{(l)}=-J_2\gamma_2^{(l-1)}-J_3\gamma_4^{(l-1)},\hskip0.825cm\gamma_4^{(l)}=J_3\gamma_3^{(l-1)}+J_4\gamma_5^{(l-1)},\\
 &\gamma_5^{(l)}=-J_4\gamma_4^{(l-1)}\hskip6.725cm
\end{eqnarray}
with $\gamma_j^{(0)}=0$ ($1$) for $j\ne5$ ($j=5$). This analysis can be summarized in the oriented graph in Fig.~\ref{fig:graph}. Each node corresponds to an operator involved in the decomposition, while an edge connect a node to the operator resulting from its commutator with the Hamiltonian. We show the corresponding coefficient and an outgoing (ingoing) edge corresponds to a $+$ (-) sign. The factor $2i$ has been omitted for the sake of simplicity.
The recurrence formulas can be easily derived from this graph. For $J_k=J\sqrt{k(5-k)}$, the spin chain becomes the one in Ref.~\refcite{cambridge} and the recurrence formulas give $\gamma_1(t)=\sin^4(2Jt)$, $\gamma_2(t)=-2\cos(2Jt)\sin^3(2Jt)$, $\gamma_3(t)=-\sqrt{(3/8)}\sin^2(4Jt)$, $\gamma_4(t)=2\cos^3(2t)\sin(2Jt)$, and $\gamma_5(t)=\cos^4(2Jt)$. At $t^*=\pi/(4J)$ we have $\gamma_1(t^*)=1$ and $\gamma_2(t^*)=\gamma_3(t^*)=\gamma_4(t^*)=\gamma_5(t^*)=0$, which corresponds to perfect state transfer.

\begin{figure}[t]
\centerline{\psfig{figure=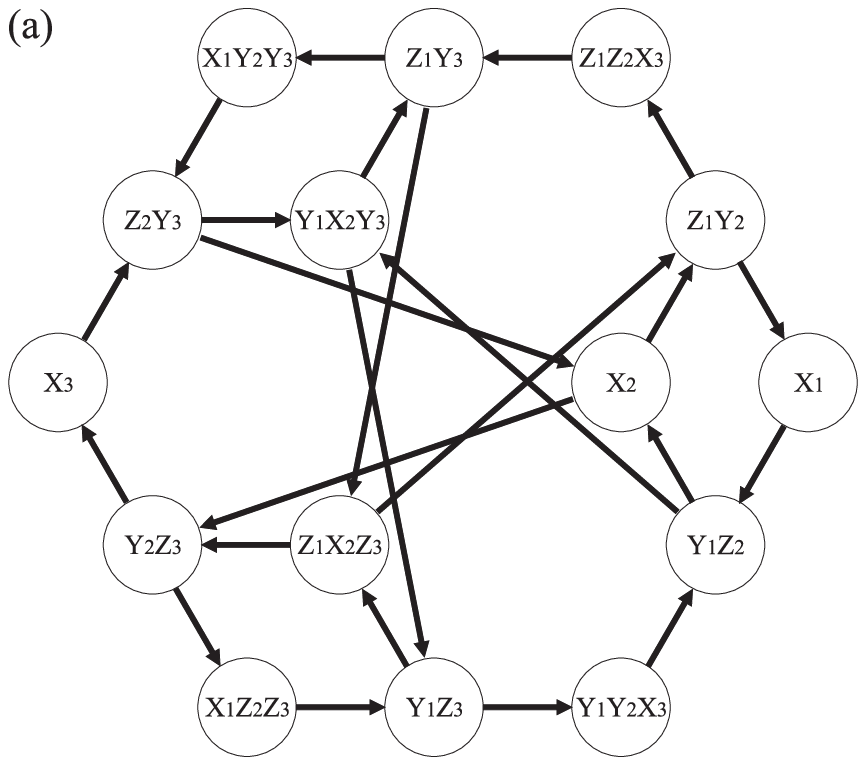,height=4.25cm}\hskip1cm\psfig{figure=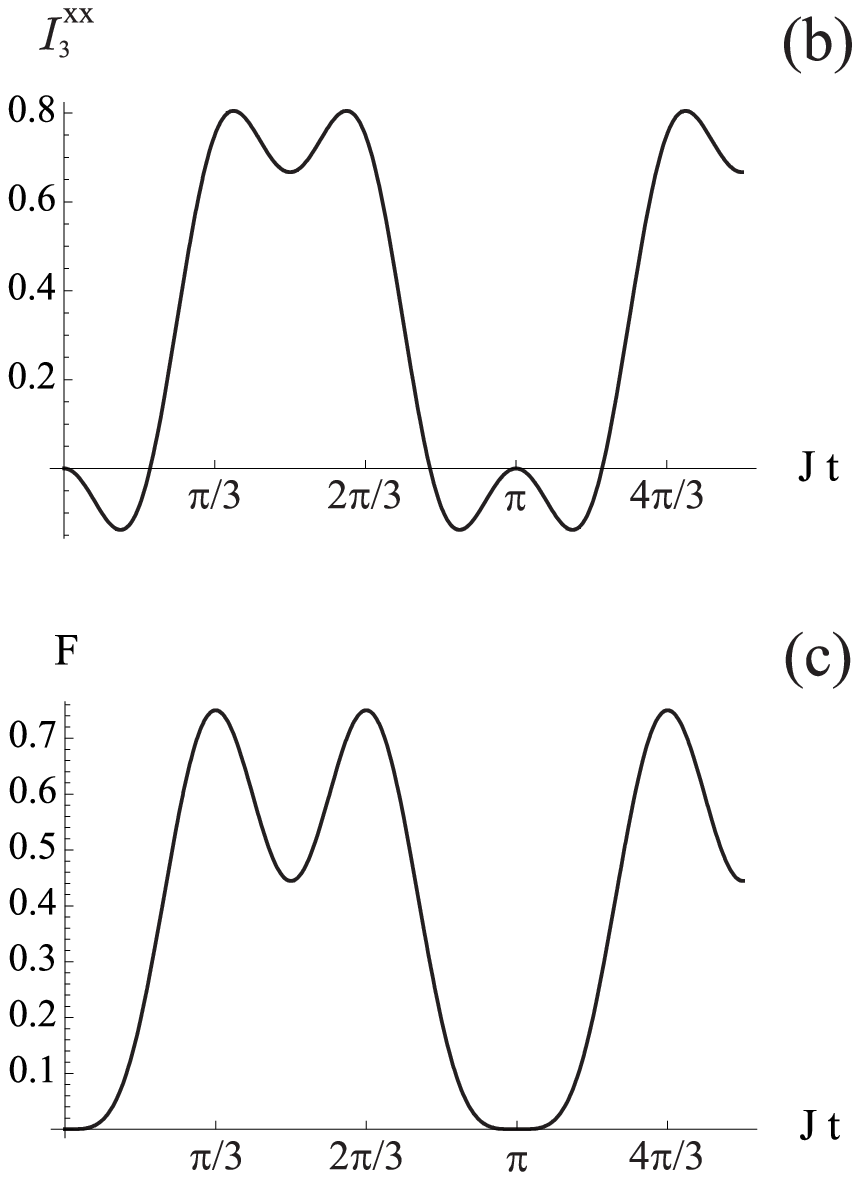,height=4.25cm}}
\caption{{\bf (a)}: Oriented graph for a 3-qubit chain with Heisenberg interaction. The coefficient $J$ above all the edges has been omitted for the sake of simplicity. {\bf (b)}: Evaluation of ${\cal I}^{XX}_{3}(t)$ by means of recurrance formulas. {\bf (c)}: State fidelity $F$ in the worst transmission case.}
\label{fig:graph2}
\end{figure}

Of course, not all the spin chains are associated to a graph with a linear structure. For example, if we analyze the Hamiltonian ${\cal\hat{H}}=\sum_{i=1}^2J(\hat{X}_i\hat{X}_{i+1}+\hat{Y}_i\hat{Y}_{i+1}+\hat{Z}_i\hat{Z}_{i+1})$, we obtain the graph in Fig.~\ref{fig:graph2} {\bf (a)}.
In this case, the decomposition of $\hat{\tilde{X}}_3$ involves two elements in which $\hat{X}_1$ is present. Therefore, the information flux from $\hat{X}_1$ to $\hat{X}_3$ (this time again we consider the initial state $\ket{\psi_0}_{23}=\ket{00}_{23}$) is ${\cal I}^{XX}_{3}(t)=\delta_1(t)\,_{23}\minisand{00}{\hat{I}_2\hat{I}_3}{00}_{23}+\delta_2(t)\,_{23}\minisand{00}{\hat{Z}_2\hat{Z}_3}{00}_{23}=\delta_1(t)+\delta_2(t)$, where $\delta_1(t)$ and $\delta_2(t)$ are the coefficients of $\hat{X}_1$ and $\hat{X}_1\hat{Z}_2\hat{Z}_3$, respectively. A plot of ${\cal I}^{XX}_{3}(t)$, evaluated by means of recurrence formulas, is presented in Fig.~\ref{fig:graph2} {\bf (b)}. In this case ${\cal I}^{XX}_{3}(t)$ never reaches $1$ (it is well known that this Hamiltonian does not allow perfect state transfer). To highlight the fact that the information flux can be used as a proper figure of merit in quantum state transfer processes, we have also plotted the state fidelity $F$ in the worst case (i.e., the transmission of the initial state $\ket{1}$) in Fig.~\ref{fig:graph2} {\bf (c)}:  The maxima in the two plots appear for the same values of scaled time $Jt$, showing the useful correspondence between information flux and state fidelity.

\section{Remarks}
\label{remarks}
We have shown how the information flux approach helps in the analysis of many-body systems, particularly in the case of spin chains. We have presented a graphical method that is useful to derive the recurrence formulas to evaluate the information flux, when it is not possible to obtain it analytically. The analogy between the information flux in the model of Ref.~\refcite{cambridge} and in a transverse Ising model with local magnetic fields has paved the way to the idea presented in a recent paper.\cite{dimitris} The linear structure of the graphs associated with the two problems allowed us to obtain a coupling-strength pattern that guarantees a perfect state transfer also in a spin-non-preserving chain.

\section*{Acknowledgments}

We thank M. S. Kim for discussions. We acknowledge financial support from UK EPSRC and QIP IRC. G.M.P. acknowledges support under PRIN 2006 ``Quantum noise in mesoscopic systems'' and under CORI 2006. M.P. is supported by The Leverhulme Trust (Grant No. ECF/40157).

\end{document}